\documentclass[preprint]{aastex}

\newcommand{\bpic}{$\beta$\,Pictoris}

\slugcomment{Draft}

\shorttitle{Discovery of an edge-on debris disk around the HD\,110058}
\shortauthors{Kasper et al.}

\begin{document}


\title{Discovery of an Edge-on Debris Disk with a Dust Ring and an Outer Disk Wing-tilt Asymmetry}


\author{Markus Kasper}
\affil{European Southern Observatory, Karl-Schwarzschild-Str. 2,
    85748 Garching, Germany}
\email{mkasper@eso.org}

\author{D\'aniel Apai\altaffilmark{1,2} }
\affil{Steward Observatory, The University of Arizona, 933 N. Cherry Avenue, Tucson AZ 85718, USA}

\author{Kevin Wagner\altaffilmark{3}}
\affil{Steward Observatory, The University of Arizona, 933 N. Cherry Avenue, Tucson AZ 85718, USA}

\and

\author{Massimo Robberto}
\affil{Space Telescope Science Institute, 3700 San Martin Dr., Baltimore MD 21218, USA}


\altaffiltext{1}{Lunar and Planetary Laboratory, The University of Arizona, 1640 E. University Blvd., Tucson, AZ 85718, USA}
\altaffiltext{2}{Earths in Other Solar Systems Team, NASA Nexus for Exoplanet System Science}
\altaffiltext{3}{The National Science Foundation, 4201 Wilson Boulevard, Arlington, Virginia 22230, USA}


\begin{abstract}
Using VLT/SPHERE near-infrared dual-band imaging and integral field spectroscopy we discovered an edge-on debris disk around the 17\,Myr old A-type member of the Scorpius-Centaurus OB association HD\,110058.
The edge-on disk can be traced to about 0$\farcs$6 or 65\,AU projected separation. In its northern and southern wings, the disk shows at all wavelengths two prominent, bright and symmetrically placed knots at 0$\farcs$3 or 32\,AU from the star. We interpret these knots as a ring of planetesimals whose collisions may produce most of the dust observed in the disk. We find no evidence for a bow in the disk, but we identify a pair of symmetric, hook-like features in both wings. Based on similar features in the Beta Pictoris disk we propose that this wing-tilt asymmetry traces either an outer planetesimal belt that is inclined with respect to the disk midplane or radiation-pressure-driven dust blown out from a yet unseen, inner belt which is inclined with respect to the disk midplane. The misaligned inner or outer disk may be a result of interaction with a yet unseen planet. Overall, the disk geometry resembles the nearby disk around Beta Pictoris, albeit seen at smaller radial scales.
\end{abstract}





\section{Introduction}

The architecture of planetary systems around young A-type stars is of particular interest since the discovery of an unexpectedly large number of large-separation giant planets (HR8799bcde: \citealt[][]{Mar08,Mar10}; \bpic\,b: \citealt[][]{Lag10}; HD106906\,b: \citealt[][]{Bai14}). In such systems, observations of spatially resolved debris disks are particularly powerful as direct probes of the location of dust belts and, therefore, the underlying planetesimal population. As the locations of the planetesimal belts, the orbits of the individual planetesimals, and the rate of dust released during their collisions are influenced by the planets in the systems, observations of debris disks provide important insights into the spatial and dynamical structure of exoplanetary systems \citep[e.g.][]{Mey07,Wya08}, including planets otherwise not detectable \citep{nes15}.

High-contrast imaging has led to high-quality images of multiple well-resolved debris disks around a handful of young A-stars (\object{HR 4796}, \object{Beta Pictoris}, \object{Fomalhaut}, \object{HD\,106906}). Somewhat unexpectedly, the dust distribution in many of these systems is concentrated in narrow dust rings, which may be tracing a single, large-separation planetesimal belt. For example, dust in the HR\,4796 system is confined to a narrow, 70\,AU ring \citep[][]{schnei99}. Similarly, the Fomalhaut system is also characterized by a $\sim$110\,AU radius dust ring with sharp boundaries, whose geometric center is also offset from the host star \citep[][]{kal05}. In the \bpic{} system, seen nearly edge-on, dust is produced in the main planetesimal belt with a radius of $\sim$65\,AU, and then the small dust grain population is driven to larger separations by radiation pressure (for a recent review of the disk structure see \citealt[][]{apa15}). Long-wavelength infrared observations and modeling of several other A-star debris disks show or suggest single or double planetesimal belts \citep[][]{su13}. Thus, well-defined, large-separation planetesimal belts appear to be common features of the brightest A-star debris disks, and some of these belts coexist with directly imaged giant planets.

Although planets have been observed in several of these systems, there is yet only one case -- \bpic{} -- where the details of the disk structure have been explained by the gravitational influence of a giant planet on the disk (see e.g. \citealt{Mou97,Aug01}). In this well-studied case, the inner disk is misaligned with the outer disk, due to the influence of the giant planet \bpic\,b \citep[][]{apa15}, which is itself on a slightly inclined orbit \citep[][]{lag12}. While bright dust belts in massive debris disks and directly imaged giant planets appear to be relatively common for around A-stars, the so far limited sample of directly imaged planets in directly imaged disks does not yet allow us to establish a firm relationship between the presence of a planet and the disk structure.

In this letter we present the discovery of a new directly imaged and well-resolved debris disk around a young A--star with an age very similar to \bpic. Our images reveal a disk structure that is dominated by a prominent debris ring as well as a misaligned outer disk, features that strongly resemble the disk of \bpic. 

Our target is \object{HD 110058} (HIP\,61782), an A0V star member of the Lower--Centaurus--Crux (LCC) subgroup of the Scorpius-Centaurus OB association \citep{dezeeuw99,riz11}. HD\,110058's revised Hipparcos parallax from van Leeuwen (2007) of $9.31\pm0.78$ mas translates to a distance of $107^{+10}_{-8}$ pc, i.e., 1\arcsec{} corresponds to about 107\,AU. The tabulated apparent magnitudes of HD\,110058 are V$=7.97$ \citep{hog00}, J$=7.64$, $H=7.59$, and K$=7.58$ \citep{cut03}. The star is situated near l$=301$\degr{} and b$=13.6$\degr{}, in the northern part of LCC, which appears to be the oldest part of the subgroup \citep{pec12} with an age of about 17 Myr \citep{chen14}. HD\,110058 has recently been observed with VLT/NACO's apodized phase plate (APP) coronagraph in L'-band \citep{mes15} and during the NICI campaign in H-band \citep{wah13}. No companion candidate or disk were detected in these data.

Nearly twenty years ago a probable debris disk was suggested around HD\,110058 to explain the unresolved infrared excess detected with the IRAS satellite \citep{mab98}. Follow-up observations with the higher-resolution ISOPHOT instrument on-board the ISO satellite \citep{moo06} confirmed that the excess emerges from a disk, based on the matching positions of star and unresolved infrared excess. These authors also estimated a fractional disk luminosity $f_d = 1.89 \times 10^{-3}$, very similar in fractional luminosity to \bpic{} \citep{lag00}. Spitzer/IRS spectra of the disk show no evidence for spectral features, such as silicate grains \citep{chen06}. Combining Spitzer with longer-wavelength MIPS data and fitting a single-temperature blackbody, \citet{chen12} found that a dust mass of $\geq0.05$ M$_\textrm{Moon}$ at $\sim25$ AU radius can explain the infrared excess. Although this dust mass is about a 100x lower than that estimated for \bpic{} \citep{den14}, the fractional infrared luminosity of the disk $1.5\times10^{-3}$ measured by \citet{chen12} is the second largest among the LCC sample stars. Only \object{HD\,100453} has a higher excess, although by a large margin with fractional luminosity of $3.9\times10^{-2}$ \citep{chen12}, and presents a luminous face-on spiral recently discovered by our team \citep{wag15}.

\section{Observations}

We observed HD\,110058 on 4 April 2015 using the IRDIFS extended mode of SPHERE \citep{beu08, kas12} under ESO Program 095.C-0389 (PI: D. Apai), a survey of 73 young A-stars in the nearby Scorpius-Centaurus OB association that will directly determine the frequency of HR\,8799b analogs. The extended mode simultaneously observes with the integral field spectrograph \citep[IFS,][]{cla08} from Y- to H-band (R$\sim$30), while the infrared dual-band imager and spectrograph \citep[IRDIS,][]{doh08} observes in two K-band filters, K1 and K2, with spectral bandwidths of about 100\,nm and central wavelengths of 2110 and 2251\,nm, respectively. The extreme Adaptive Optics system SAXO \citep{fus14} corrected for atmospheric turbulence, and an apodized Lyot coronagraph, optimized for observations from Y- to H-band with an inner working angle defined by the mask diameter of 185\,mas, reduced diffraction residuals.

We acquired datacubes with a total exposure time of 1600 seconds for each instrument. The field rotation exploitable for angular differential imaging \citep[ADI,][]{mar06} varied by 14.8\degr{} during the observation. In addition to standard calibrations (sky, flat field, etc.), we recorded i) coronagraphic images with four centrally symmetric satellite speckles (created by generating sine aberrations with the adaptive optics' deformable mirror) to determine the star's position behind the mask, and ii) images with the star shifted away from behind the Lyot mask for flux calibration. For the latter, a neutral density filter with factor ten attenuation (ND1) was inserted to avoid PSF saturation. The SPHERE webpage (http://www.eso.org/sci/facilities/paranal/instruments/sphere.html) provides the filter transmission curves. Following this standard observing procedure of SPHERE, we calibrated photometry and astrometry with high fidelity. SPHERE frequently observes astrometric calibration fields to monitor plate scale and field orientation. The IRDIS pixel size  $12.251\pm0.005$\,mas and the IFS images plate scale $7.46\pm0.02$\,mas per pixel are those provided in the SPHERE user manual.

We used the SPHERE data reduction pipeline \citep{pav08} to create backgrounds, bad pixel maps and flat fields. We reduced the raw data by subtracting the background, replacing bad pixels by the median of the nearest valid pixels, and finally dividing by the flat field. We also used the pipeline to create the IFS x-y-$\lambda$ data cube. Parts of this cube were collapsed along the wavelength axis to create broad-band images in Y-band (990-1100\,nm), J-band (1140-1350\,nm) and the short part of H-band (1490-1640\,nm, a filter in SPHERE cuts off the long-end of the H-band in front of the IFS in order to limit sky background).

We centered the images using the satellite spots mentioned above and removed large-scale structure in the images (such as the residual halo after AO correction) through spatial high-pass filtering. The smoothing length for the spatial filtering is a free parameter, adjusted to remove as much of the halo as possible while leaving astronomical features intact. We carefully verified that the edge-on disk morphology is not affected by this processing step. Finally, for each individual exposure, we performed a principal component analysis \citep[e.g.][]{ama12} using other exposures, which differed in field rotation by at least 7.4\degr{}, half of the total field rotation. Conservatively, we used only the first principal component as a PSF estimate and subtracted it from the original frame. Since the relatively small amount of total field rotation leads to a certain amount of self-subtraction of a point source (size $\lambda/D$) at angular separations between 200\,mas (Y-band) and 420\,mas (K-band), one must be cautious when trying to determine the precise morphology of structures at smaller angular separations.

\section{Results}

\begin{figure}
\includegraphics[angle=0,scale=0.5]{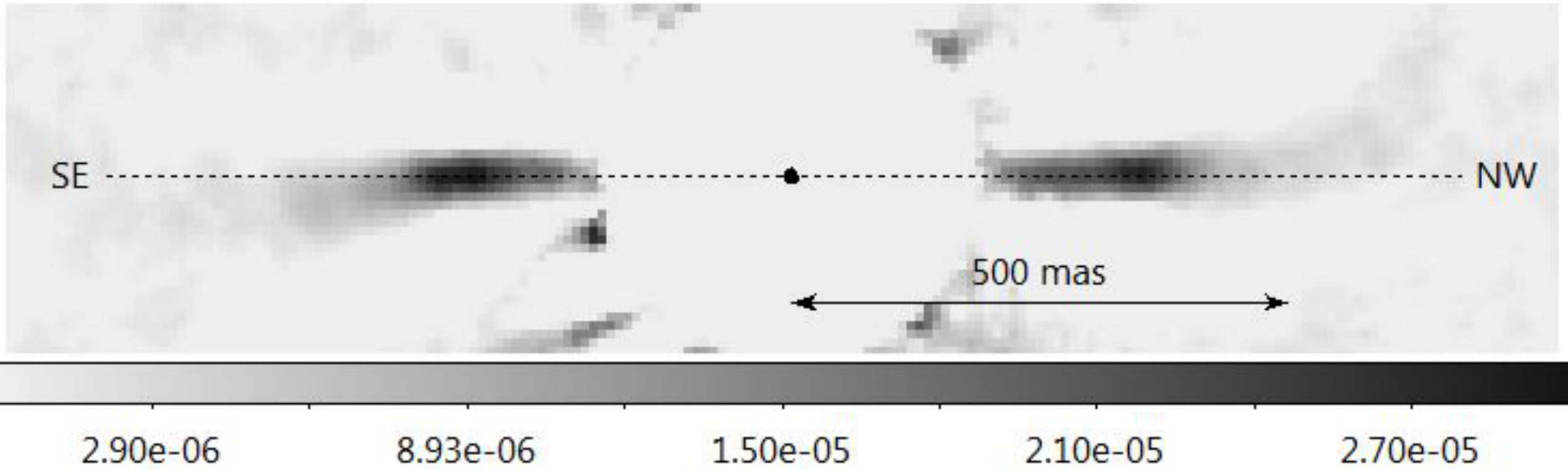}
\caption{The edge-on disk of HD 110058 in YJH-band. The image is rotated by 295\degr{} degrees from North up and East left to align the disk with the horizontal axis. Y-, J-, and H images have been normalized individually to the maximum pixel intensity of the non-coronagraphic stellar PSF and averaged. Hence, the scale represents the YJH average intensity contrast with respect to the star. The stretch is linear. The star's position is indicated in the center.}
\label{fig1}
\end{figure}

\begin{figure}
\includegraphics[angle=0,scale=0.3]{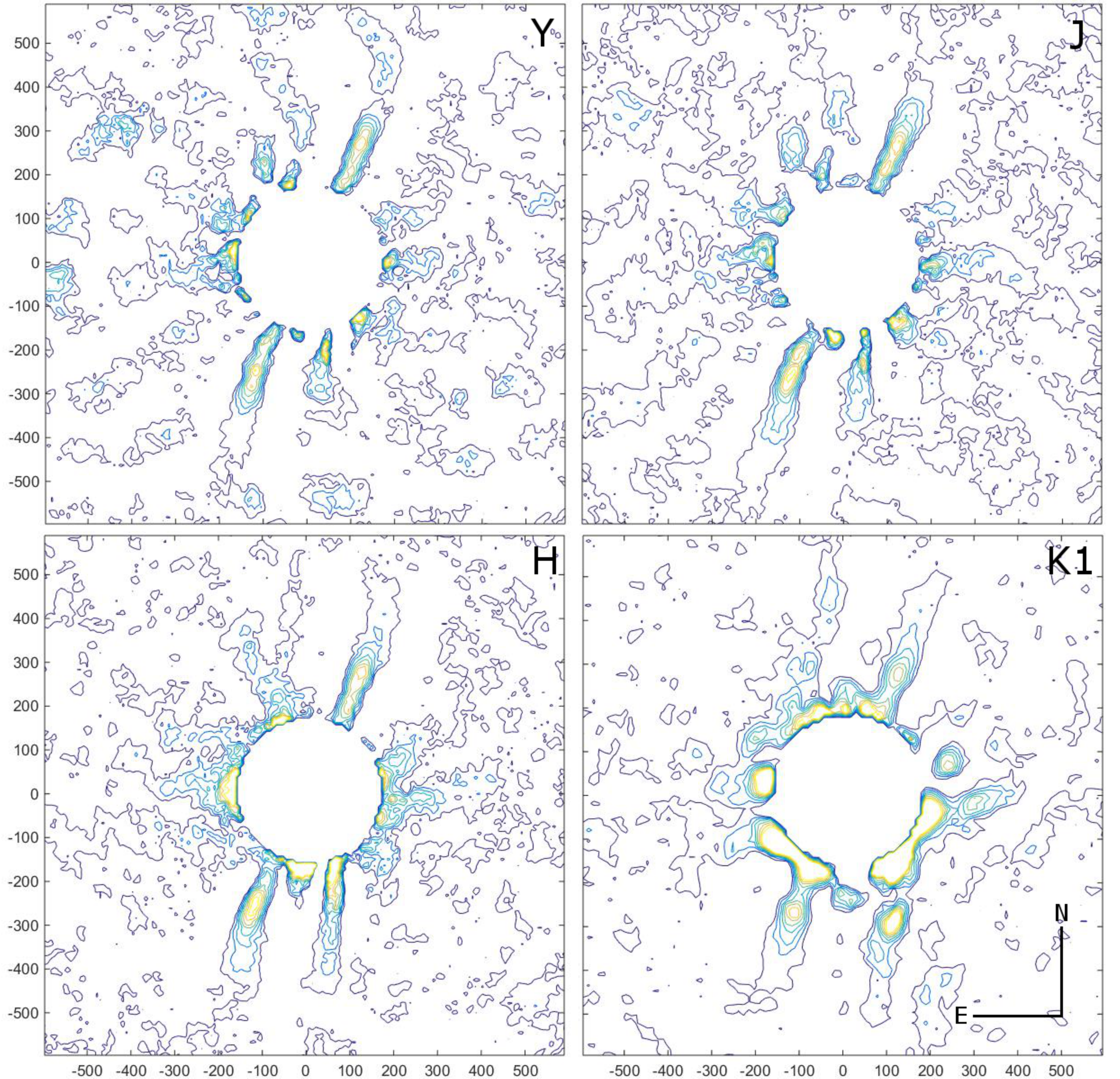}
\caption{Contour plots (eight linearly spaced levels between 0 and the maximum pixel intensity of the disk, brighter image artefacts are saturated) of HD\,110058 in Y, J, H, K1-bands (upper left to lower right). The coordinates are given in milliarcseconds relative to the star.}
\label{fig2}
\end{figure}

The SPHERE data reveal the presence of an edge-on debris disk around HD\,110058. Figure\,\ref{fig1} shows the IFS image where all the wavelengths channels of Y, J, and H-band have been co-added. The disk is seen nearly edge-on with a position angle $155\pm1$\degr{}, i.e., the disk is oriented from the Southeast to the Northwest. The disk possesses two symmetric areas of enhanced surface brightness visible in all bands. This would be expected for a ring-like disk, optically thin and seen edge-on. The radial distance of these knots, i.e., the radius of the ring is about 300\,mas or 32\,AU at the distance of 107\,pc. The knots are best defined in K-band (see lower right panel of Figure\,\ref{fig2}) and symmetric around the star within the astrometric accuracy of the data of better than one pixel, i.e., better than about 10\,mas. This suggests a low eccentricity of the ring $e<0.035$, if the disk would be observed along its semi-minor axis excluding an eccentricity similar to the one of the Fomalhaut disk. If viewed along the semi-major axis, the knots would be symmetric even for high eccentricity.

Beyond this separation, the surface brightness drops rapidly and can no longer be detected in our data at about 600\,mas or 65\,AU. The other structures near the masked central area are PSF residuals. In particular, the nearly vertical streak in the H-band image (see lower left panel of Figure\,\ref{fig2}) starting slightly West of the center and extending towards the South is diffraction from the telescope secondary's support structure which does not fully subtract out by the ADI reduction.

The disk is not spatially resolved in the direction perpendicular to the disk plane. The disk thickness corresponds to the PSF's full width at half maximum (FWHM), which is of the order of 40\,mas in Y- and J-bands, 45\,mas in H-band, and 60\,mas in K-band, i.e., 4-6\,AU at the distance of 107\,pc, which argues for a disk scale height less than $\sim4$\,AU and a near-perfect edge-on viewing angle. Furthermore, the star lies on the disk mid-plane with an accuracy better than one pixel (7.46\,mas or 0.8\,AU). There is no sign of curvature of the disk's projected image. 

There is, however, clear evidence for counter-clockwise warps on both sides of the outer disk, starting at the radial distance of the ring. To the Southeast the disk curves away by about 15\degr{}. At low SNR, the images apparently show a similar warp to the Northwest resembling a hook ending almost perpendicular to the diskplane. Figure\,\ref{fig2} shows the disk image in the individual bands as observed by the IFS (Y, J, H) and IRDIS (K1). The edge-on disk is easily seen at all wavelengths extending from the Southeast to the Northwest. The Northwest extended hook feature is most prominent in Y-band. At longer wavelengths, in J- and H-band, we see the inner warp, but the outer parts of the hook-like features are no longer detected. We nevertheless believe that this feature is real, because it connects to the inner warp seen at all wavelengths. The feature also robustly appears when combining randomized non-overlapping subsets of our data, so it is unlikely to be produced by intermittent image artifacts. 

The ratio of the integrated flux from the visible parts of the disk starting at 165 mas (limited by PSF residuals and the Lyot mask further in) to the stellar flux amounts to $1.8\pm0.12\times10^{-4}$, $1.8\pm0.1\times10^{-4}$, $2.2\pm0.08\times10^{-4}$, and $1.8\pm0.15\times10^{-4}$ in the Y, J, H, and K1 bands, respectively. The fractional near-infrared luminosity of the HD\,110058's disk with respect to the star is therefore almost gray and about ten times lower than the ones of HR\,4796 \citep{schnei99}, HD\,115600 \citep{cur15}, or \bpic{} \citep{lag00}.

\begin{figure}
\includegraphics[angle=0,scale=0.5]{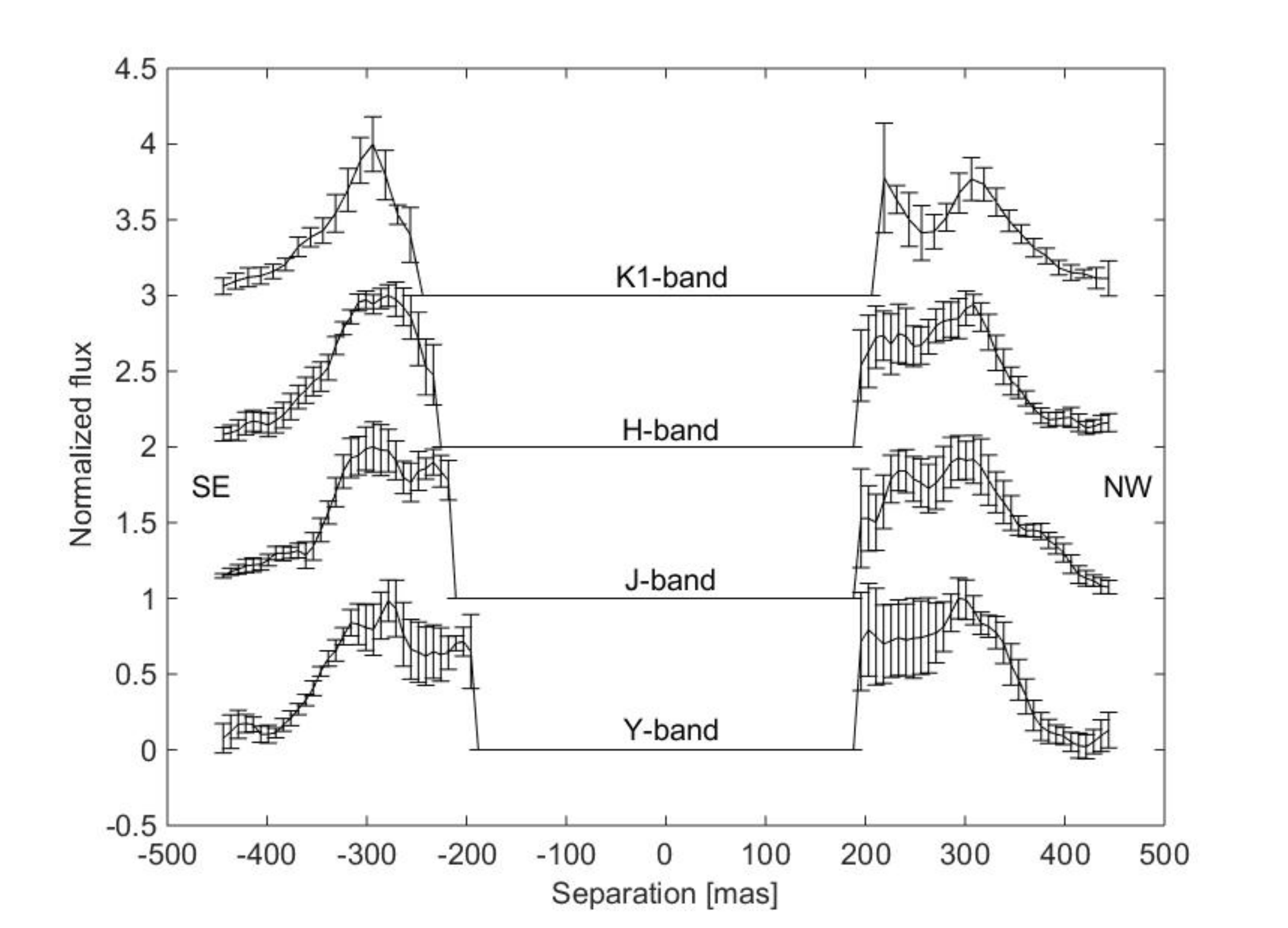}
\caption{Radial profiles of the brightness of the edge-on disk in the Y, J, H, and K1 bands. The profiles were normalized to their peak brightness and vertically shifted.}
\label{fig3}
\end{figure}

Figure\,\ref{fig3} plots the radial surface brightness profiles integrated over one PSF FWHM perpendicular to the disk plane. The error bars represent the standard deviation of flux measured in a similar manner along other position angles not aligned with the disk. Comparing the different wavebands, there appears to be two trends with wavelength: the knot indicating the ring moves slightly outwards with wavelength, and it becomes sharper at the inner edge. Both effects are expected for an optically thin disk as optical depth decreases with wavelength enhancing parts with higher optical depth at longer wavelengths.

By inserting artificial planets in the data, we establish detection limits around 13\,mag contrast in the combined YJH-band image and 12\,mag contrast in K-band at separations larger than about 0$\farcs$3. The radial contrast beyond these separations is essentially flat and limited by sky background and detector noise. Close to our inner working angle of about 0$\farcs$2, or right in the plane of the disk, the detection contrast is reduced by 0.5-1 magnitudes. At the distance of 107\,pc and assuming an age of 17\,Myr, we compare our sensitivity limits to the NIR brightnesses of planets predicted by the hot start models of \citet{bar03}. This model comparison suggests that our sensitivity is sufficient to detect $\sim$3-4\,M$_\textrm{Jup}$ in our field of view. The non-detection in our current data thus argues against the presence of a super-Jupiter comparable to HR\,8799\,bcde or \bpic{}\,b at separations larger than $\sim$25\,AU from HD\,110058. Nevertheless, we point out that a \bpic{} analog ($\sim$8\,M$_\textrm{Jup}$ on a $\sim$8\,AU semi-major axis orbit) would remain undetectable on our images.

\section{A Ring and Outer Disk Wing-tilt Asymmetry}
In the following we briefly discuss the disk geometry and place it the context of directly imaged A-star debris disks and exoplanetary systems. We can trace the disk in scattered light between 0$\farcs$2 and $\sim$0$\farcs$6, corresponding to projected separations of 20\,AU to 65\,AU. The outer radius observed is limited by the sensitivity of our relatively shallow observations and the true radius is likely to be significantly larger. 

The most important features visible on our images are the northern and southern 'knots' at $\sim$32\,AU and the two "hooks", counter-clock-wise deviations in the disk structure at radii beyond $\sim$40\,AU.  
The fact that our observations show no evidence for a disk scaleheight comparable to the resolution of our observations ($\sim$4\,AU) demonstrates that the HD\,110058 disk is both geometrically flat and seen nearly exactly edge-on. The fact that the disk is flat suggests that is depleted in gas, otherwise the small dust grains traced in scattered light would couple to the gas, leading to flared disks with scale heights comparable to the orbital radius typical to protoplanetary disks.

A straightforward interpretation for the origin of the symmetric knots is the presence of a dust belt with radius $\sim$32\,AU around HD\,110058. Such ring structures have been seen in several other bright, directly imaged inclined A-star debris disks, with the $\sim$70\,AU radius dust ring around HR\,4796 \citep{schnei99} and the $\sim$110\,AU ring around Fomalhaut \citep{kal05} being the best known examples. Also the recently discovered disk around the young F-star HD\,115600 \citep{cur15}, another member of the Sco-Cen OB Association, clearly shows a ring-like morphology. Finally, the edge-on disk around the young A-star \bpic{} hosts a prominent dust belt at $\sim$65\,AU. In all of these disks the dust belts trace a planetesimal belt fueling a collisional cascade, which in turn replenishes the fine dust in the system. 

The other conspicuous structures seen in the HD\,110058 disk are the two "hooks", one at each side of the disk. These appear to emerge from the disk midplane at the planetesimal belt radius and extend to at least $\sim$65\,AU from the star and $\sim$17\,AU above the disk midplane. We interpret these wing tilt asymmetries as a major warp in the outer disk or, in other words, an outer disk misaligned from the inner disk.

While such wing tilt asymmetries are difficult to see in nearly face-on viewing geometries, a similar asymmetry has been documented in the edge-on \bpic{} disk. The \bpic{} inner disk is misaligned by $\sim4^\circ$ with respect to the planetesimal ring at $\sim$65\,AU \citep[][]{apa15}. This feature has been explained by the precession of the planetesimal orbits due to the gravitational influence of \bpic\,b \citep[e.g.][]{Mou97, Aug01, nes15}, a $\sim13$\,M$_\textrm{Jup}$ giant planet orbiting at an $\sim$8\,AU orbit inclined at $\sim$1.8\degr{} \citep{Lag10}. The fine dust produced in the inner, inclined dust belt is thus blown out to large orbital radii and forms an asymmetric hook-like feature identified in \bpic{} as the "butterfly asymmetry" \citep[][]{kal95}. \citet{apa15} summarize the asymmetries in the \bpic{} disk and show that all axisymmetric structures are, in fact, consistent with disk structures predicted by models that consider the planets influence and radiation pressure. 

Similarly to \bpic{}, the HD\,110058 disk displays prominent misalignments between the medium-separation planetesimal belt and the outer disk. We speculate that the misaligned outer disk in HD\,110058 may arise, similarly to \bpic, due to a misaligned inner disk. This would suggest the presence of a misaligned, inner planet orbiting at about half the observed angular extent of the outer disk mis-alignment \citep{daw11}. Alternatively, HD\,110058 may also be the result of the interaction of the outer disk and a very large-separation ($>70$\,AU), yet unseen planet, with a warp preceding from inward, rather than outward as in the case of \bpic. Detailed numerical simulations and deeper high-contrast imaging observations can discriminate between these possibilities and test the presence and location of a giant planet on inclined orbit around HD\,110058.

The edge-on disk in HD\,110058 complements the other large directly imaged A-star debris disks (\bpic, HR\,4796, Fomalhaut, HD\,106906). It suggests that single or double planetesimal belts are very common in these disks and that mis-aligned inner and outer disks occur in a significant fraction of A-star disks. 

\subsection{Conclusions}

We present here the discovery of an edge-on debris disk around a young A--star with an age similar to \bpic. We trace scattered light from the disk between 0$\farcs$2 and 0$\farcs$6, corresponding to about 20 to 65\,AU in the system. The disk shows two prominent, symmetrically placed bright knots, which we interpret as projections of a single, relatively massive planetesimal belt with a radius of $\sim$32\,AU. Furthermore, we identify two hook-like features, i.e., large separation counter-clock-wise wing-tilt asymmetries in the disk. We propose that, similarly to \bpic, the wing-tilt asymmetry may result from dust that is blown out by radiation pressure from an inner, inclined disk. Alternatively, the outer disk may be shaped by a yet unseen planet on an inclined orbit.

The bright, well-resolved edge-on disk in HD\,110058 adds to the important but small sample of directly imaged debris disks around A-type stars, some of which also hosts directly imaged planets. If a planet would be detected in the system, HD\,110058 would provide a novel target in which disk-planet interactions could be tested in a system similar to \bpic.



\acknowledgments
This work is based on observations performed with VLT/SPHERE under program ID 095.C-0389A and supported by the National Science Foundation Graduate Research Fellowship under Grant No. 2015209499. The results reported herein benefited from collaborations and/or information exchange within NASA’s Nexus for Exoplanet System Science (NExSS) research coordination network sponsored by NASA’s Science Mission Directorate. We would also like to thank the astronomers and the instrument support team at the VLT for the observations in service mode.



{\it Facilities:} \facility{VLT (SPHERE)}.

\clearpage



\begin{thebibliography}{}
		\bibitem[Apai et al.(2015)]{apa15} Apai, D., Schneider, G., Grady, C.~A., et al., 2015, \apj, 800, 136
		\bibitem[Amara \& Quanz(2012)]{ama12} Amara, A., Quanz, S.~P., 2012, \mnras, 427, 948
		\bibitem[Augereau et al.(2001)]{Aug01} Augereau, J.~C., Nelson, R.~P., Lagrange, A.~M., Papaloizou, J.~C.~B., Mouillet, D., 2001, \aap, 370, 447
		\bibitem[Baraffe et al.(2003)]{bar03} Baraffe, I., Chabrier, G., Barman, T.~S., Allard, F., Hauschildt, P.~H., 2003, \aap, 402, 701
		\bibitem[Bailey et al.(2014)]{Bai14} Bailey, V., Meshkat, T., Reiter, M., et al., 2014, \apjl, 780, L4
		\bibitem[Beuzit et al.(2008)]{beu08} Beuzit, J.-L., Feldt, M., Dohlen, K., et al., 2008, SPIE Conference Series, Vol. 7014
		\bibitem[Chen et al.(2006)]{chen06} Chen, C. H., Sargent, B. A., Bohac, C., et al., 2006, \apjs, 166, 351
		\bibitem[Chen et al.(2012)]{chen12} Chen, C. H., Pecaut, M., Mamajek, E. E., Su, K. Y. L., Bitner, M., 2012, \apj, 756, 133
		\bibitem[Chen et al.(2014)]{chen14} Chen, C. H., Mittal, T., Kuchner, M., et al., 2014, \apjs, 211, 25
		\bibitem[Claudi et al.(2008)]{cla08} Claudi, R. U., Turatto, M., Gratton, R. G., et al., 2008, SPIE Conference Series, Vol. 7014
		\bibitem[Currie et al.(2015)]{cur15} Currie, T., Lisse, C.~M., Kuchner, M., et al., 2015, \apjl, 807, L7
		\bibitem[Cutri et al.(2003)]{cut03} Cutri, R. M., Skrutskie, M. F., van Dyk, S., et al., 2003, yCat, 2246, 0
		\bibitem[Dawson et a.(2011)]{daw11} Dawson, R.~I., Murray-Clay, R.~A., Fabrycky, D.~C., 2011, \apjl, 743, L17
		\bibitem[Dent et a.(2014)]{den14} Dent, W.~R.~F., Wyatt, M.~C., Roberge, A., et al., 2014, Science, 343, 1490
		\bibitem[de Zeeuw et al.(1999)]{dezeeuw99} de Zeeuw, P. T., Hoogerwerf, R., de Bruijne, J. H. J., Brown, A. G. A., Blaauw, A. , 1999, \aj, 117, 354
		\bibitem[Dohlen et al.(2008)]{doh08} Dohlen, K., Langlois, M., Saisse, M., et al., 2008, SPIE Conference Series, Vol. 7014
		\bibitem[Fusco et al.(2014)]{fus14} Fusco, T., Sauvage, J.-F., Petit, C., et al., 2014, SPIE Conference Series, Vol. 9148
		\bibitem[Hog et al.(2000)]{hog00} Hog, E., Fabricius, C., Makarov, V. V., et al., 2000, \aap, 355, L27
		\bibitem[Kalas \& Jewitt(1995)]{kal95} Kalas, P., Jewitt, D., 1995, \aj, 110, 794
		\bibitem[Kalas et al.(2005)]{kal05} Kalas, P., Graham, J.~R., Clampin, M., 2005, \nat, 435, 1067
		\bibitem[Kasper et al.(2012)]{kas12} Kasper, M., Beuzit, J.-L., Feldt, M., et al., 2012, The Messenger, 149, 17
		\bibitem[Lagrange et al.(2000)]{lag00} Lagrange, A.-M., Backman, D.  E., Artymowicz, P., 2000, Protostars and Planets IV, ed. V. Manning, A. P. Boss, \& S. S. Russell (Tucson, AZ: Univ. Arizona Press), 639
		\bibitem[Lagrange et al.(2010)]{Lag10} Lagrange, A.-M., Bonnefoy, M., Chauvin, G., et al., 2010, Science, 329, 57
		\bibitem[Lagrange et al.(2012)]{lag12} Lagrange, A.-M., Boccaletti, A., Milli, J., et al., 2012, \aap, 542, A40
		\bibitem[Mannings \& Barlow(1998)]{mab98} Mannings, V., Barlow, M. J., 1998, \apj, 497, 330
		\bibitem[Marois et al.(2006)]{mar06} Marois, C., Lafreni\`ere, D., Doyon, R., Macintosh, B., Nadeau, D., 2006, \apj, 641, 556
		\bibitem[Marois et al.(2008)]{Mar08} Marois, C., Macintosh, B., Barman, T., et al., 2008, Science, 322, 1348
		\bibitem[Marois et al.(2010)]{Mar10} Marois, C., Zuckerman, B., Konopacky, Q.~M., Macintosh, B., Barman, T., 2010, \nat, 468, 1080
		\bibitem[Meshkat et al.(2015)]{mes15} Meshkat, T., Bailey, V. P., Su, K. Y. L, 2015, \apj, 800, 5
		\bibitem[Meyer et al.(2007)]{Mey07} Meyer, M.~R., Backman, D.~E., Weinberger, A.~J., Wyatt, M.~C, 2007, Protostars and Planets V, p573
		\bibitem[Mo\'or et al.(2006)]{moo06} Mo\'or, A., \'Abrah\'am, P., Derekas, A., et al., 2006, \apj, 644, 525
		\bibitem[Mouillet et al.(1997)]{Mou97} Mouillet, D., Larwood, J.~D., Papaloizou, J.~C.~B., Lagrange, A.~M., 1997, \mnras, 292, 896
		\bibitem[Nesvold \& Kuchner(2015)]{nes15} Nesvold, E.~R. \& Kuchner, M.~J., 2015, submitted to \apj
		\bibitem[Pavlov et al.(2008)]{pav08} Pavlov, A., M\"oller-Nilsson, O., Feldt, M., et al., 2008, SPIE Conference Series, Vol. 7019
		\bibitem[Pecaut et al.(2012)]{pec12} Pecaut, M. J., Mamajek, E. E., Bubar, E. J., 2012, \apj, 746, 154
		\bibitem[Rizzuto et al.(2011)]{riz11} Rizzuto, A. C., Ireland, M. J., Robertson, J. G., 2011, \mnras, 416, 3108
		\bibitem[Schneider et al.(1999)]{schnei99} Schneider, G., Smith, B.~A., Becklin, E.~E, et al., 1999, \apjl, 513, L127
		\bibitem[Schneider et al.(2005)]{schnei05} Schneider, G., Silverstone, M.~D., Hines, D.~C., 2005, \apjl, 629, L117
		\bibitem[Su et al.(2013)]{su13} Su, K.~Y.~L., Rieke, G.~H., Malhotra, R., et al., 2013, \apj, 763, 118
		\bibitem[Wahhaj et al.(2013)]{wah13} Wahhaj, Z., Liu, M. C., Nielsen, E. L., et al., 2013, \apj, 773, 179
		\bibitem[Wagner et al.(2015)]{wag15} Wagner, K., Apai, D., Kasper, M., Robberto, M., 2015, submitted to \apjl
		\bibitem[Wyatt et al.(2008)]{Wya08} Wyatt, M.~C., 2008, \araa, 46, 339
		\bibitem[Wyatt et al.(2010)]{Wya10} Wyatt, M.~C., Booth, M., Payne, M.~J., Churcher, L.~J., 2010, \mnras, 402, 657
\end{thebibliography}
\end{document}